\documentclass[10pt,letterpaper]{article}
\usepackage{opex3}
\usepackage[noadjust]{cite}
\usepackage{srcltx}
\usepackage{psfrag}
\usepackage{epsfig}
\usepackage{amsmath}
\usepackage{subfigure}
\usepackage{subfig}
\begin{document}

\title{Spatio-Temporal Metasurface for Real-time 2-D Spectrum Analysis}

\author{Shulabh Gupta$^1$ and Christophe Caloz$^{1,2}$}
\address{$^{1}$Department of Electrical Engineering, Polytechnique Montr\'{e}al, Qu\'{e}bec, Canada.}
\address{$^{2}$King Abdulaziz University, Jeddah, Saudi Arabia.}
\email{shulabh.gupta@polymtl.ca} 

\begin{abstract}
A spatio-temporal metasurface is proposed to decompose in real time the temporal frequencies of electromagnetic waves into spatial frequencies onto a two-dimensional plane. The metasurface is analyzed and demonstrated using Fourier analysis. The required transmittance function is derived from an equivalent free-space optical system consisting of the cascade combination of a wedge, a diffraction grating and a focusing lens. The metasurface must exhibit both multi-resonance over a broad bandwidth and 1-D grating-type scanning to achieve the specified 2-D frequency scanning in space. Compared to state-of-the art related systems, the proposed metasurface system is more compact as it requires only one dispersive structure, while maintaining the high frequency resolution that characterizes 2-D spatial-temporal mapping systems.
\end{abstract}

\ocis{(050.0050) Diffraction and gratings, (070.0070) Fourier optics and signal processing, (230.0230) Optical devices, (300.0300) Spectroscopy, (320.0320) Ultrafast optics, (050.7330) Volume gratings, (070.1170 ) Analog optical signal processing, (070.7345) Wave propagation.}

\bibliographystyle{osajnl}
\bibliography{Gupta_2DSpectrum_Analyzer_OpticsExpress_2014_References}

\section{Introduction}

Real-time spectrum analysis is a ubiquitous signal processing operation in science and engineering. It involves real-time frequency discriminating devices that separate the various spectral components of a signal in either the space domain or the time domain. Typical applications include spectral analysis for instrumentation, electromagnetics and biomedical imaging~\cite{Spatial_Disperser,Acoustooptic_SA,Gupta_TMTT_04_2009}, ultrafast optical signal processing~\cite{SA_Application1_Jalali,SA_Application2_Jalali}, and dense wavelength demultiplexing communication systems~\cite{2DWDemux_Weiner,Spatial_Disperser_Weiner}, to name a few.

The heart of such systems is a spatial temporal dispersive device. Typical devices are optical prisms, diffraction gratings, arrayed-waveguide gratings (AWGs)~\cite{Saleh_Teich_FP,Goodman_Fourier_Optics}, Bragg gratings \cite{Kashyap_FBG}, phasers \cite{Caloz_MM_2012}, leaky-wave antennas \cite{Caloz_McrawHill_2011} and virtual image phased arrays (VIPAs) \cite{VIPA_Fujitsu}. While these systems have been extensively used for frequency discrimination along 1-dimension, there has always been a huge demand to increase their frequency sensitivity and resolution by exploiting a second dimension of space.

For this reason, various 2-D real-time frequency analysis systems have been recently reported in the literature. The first system of this type was introduced in a patent filed by Dragone and Forde filed in 1999, and used an arrayed-waveguide grating, a diffraction grating and focussing lenses~\cite{Spatial_Disperser_Original_Patent}. Several related systems have been proposed since then, all utilizing free-space propagation between two diffractive elements to achieve the same effect~\cite{Spatial_Disperser,2DWDemux_Weiner,AWG_2D_SA}. However, while offering enhanced resolution and sensitivity, these 2-D spectrum analyzers are bulky systems.

This paper presents a metasurface approach of 2-D temporal to spatial frequency decomposition. Compared to the aforementioned approaches, the resulting system exhibits dramatically reduced size, since it utilizes a \emph{single electromagnetic dispersive element} instead of two. Metasurfaces generally consist of nonuniform spatial arrays of subwavelength scattering particles and provide unprecedented flexibility in controlling wavefronts in space, time or both space and time~\cite{meta1, meta2, meta3, meta4, FlatOptics_Capasso, Achromatic_MS_Capasso}.

\section{Principle of 2-D Spatial Spectral Decomposition}

\subsection{Conventional System}

\begin{figure}[htbp]
\begin{center}
\psfrag{h}[c][c][0.6]{\shortstack{incident wave\\$\psi(x,y, z; t)$}}
\psfrag{g}[c][c][0.6]{$z=0$}
\psfrag{a}[c][c][0.6]{\shortstack{spectral decomposition\\along $x-$axis}}
\psfrag{f}[c][c][0.6]{\shortstack{spectral decomposition\\along $y-$axis}}
\psfrag{d}[c][c][0.6]{\shortstack{output plane \\$[x_0(\nu),\;y_0(\nu)]$}}
\psfrag{e}[c][c][0.6]{$ f$}
\psfrag{x}[c][c][0.6]{$x$}
\psfrag{y}[c][c][0.6]{$y$}
\psfrag{z}[c][c][0.6]{$z$}
\psfrag{v}[c][c][0.6]{$\nu$}
\psfrag{r}[c][c][0.6]{focussing lens}
\psfrag{b}[c][c][0.6]{$z_2$}
\psfrag{c}[c][c][0.6]{$z_1$}
\psfrag{u}[c][c][0.6]{\shortstack{dispersive \\element \#1}}
\psfrag{n}[c][c][0.6]{\shortstack{dispersive \\element \#2}}
\psfrag{i}[c][c][0.6]{$\nu_\text{min}$}
\psfrag{j}[c][c][0.6]{$\Delta\nu_\text{FSR}\{$}
\psfrag{k}[c][c][0.6]{$\nu_\text{max}$}
\psfrag{l}[c][c][0.6]{FSR$_1$}
\psfrag{m}[c][c][0.6]{FSR$_2$}
\psfrag{q}[c][c][0.6]{FSR$_{(M-1)}$}
\psfrag{o}[c][c][0.6]{FSR$_M$}
\psfrag{p}[c][c][0.8]{$z=z_1+z_2+f$}
\psfrag{s}[c][c][0.8]{bandwidth $\Delta\nu = \nu_\text{min} - \nu_\text{max}$}
\includegraphics[width=\columnwidth]{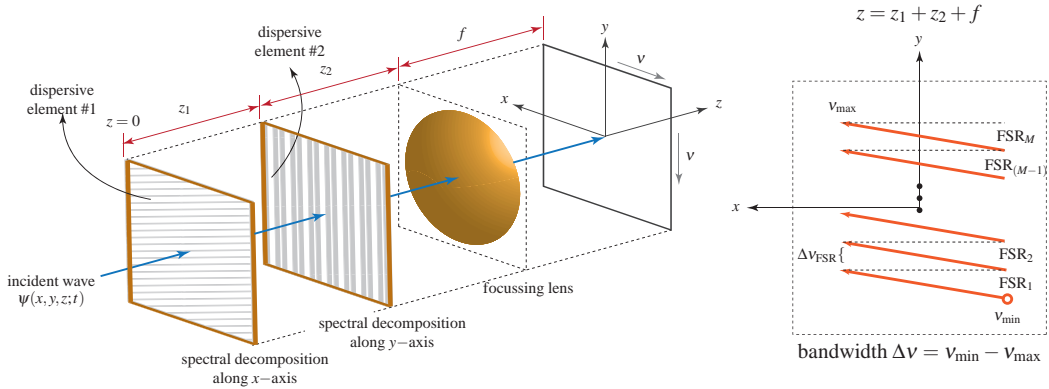}
\caption{Conventional optical system decomposing the temporal spectrum of a broadband wave, $\psi(x,y,z;t)$, in two spatial dimensions using two dispersive elements.}
\label{Fig:Conventional2D}
\end{center}
\end{figure}

Figure~\ref{Fig:Conventional2D} depicts a conventional 2-D decomposition system. This system operates as follows. Consider the modulated pulse,

\begin{equation}
\psi(x,y,z;t) = \text{Re}\{\Psi(x,y;t)e^{-j(kz - \nu_0 t)}\},
\end{equation}

\noindent where $\nu_0$ is the carrier frequency, $k = 2\pi/\lambda_0 = 2\pi \nu_0/c$ is the wavenumber with $\lambda_0$ and $\nu_0$ being the wavelength and temporal frequency, respectively, and $\Psi(x,y;t)$ is the pulse envelope, with spectral bandwidth $\Delta\nu = \nu_\text{max} - \nu_\text{min}$. At the input of the first dispersive element, the wave is $\psi(x,y,0_-;t) = \text{Re}\{\Psi(x,y;t)\}$, where the time dependence $e^{j\nu_0 t}$ has been dropped for simplicity. Dispersive element~$\#1$ spectrally decomposes $\psi(x,y,0_-; t)$ along the $x-$direction, which results in the wave $\psi(x,y,z_{1-};t)$ in the plane $z=z_{1-}$. Then dispersive element~$\#2$ decomposes this wave along the $y-$direction, which results in the wave $\psi(x,y,z_{-2};t)$ in the plane $z=z_{2-}$. This wave is finally focussed by the lens onto the focal plane $z=z_1+z_2+f$, so that each frequency $\nu$ (or wavelength $\lambda$) of the input wave is mapped onto a specific point $[x_0(\nu),\;y_0(\nu)]$ in this plane. The first dispersive element is typically either an AWG or a VIPA, with multiple free-spectral ranges~(FSR), $\Delta\nu_\text{FSR}$, within the operation bandwidth, $\Delta\nu$, of the system, so that $x_0(\nu) = x_0(\nu + n\Delta \nu_\text{FSR})$, where $n$ is an integer. This leads to the so-called \emph{spectral shower}, represented in the right of Fig.~\ref{Fig:Conventional2D}. The total depth of the system is $\Delta z = z_1 + z_2 + f$. In principle, this depth can be reduced to $\Delta z_\text{min} = (z_1 + f)$ by placing the lens directly after the second dispersive element \cite{Goodman_Fourier_Optics}, but really requires the spacing $z_1$ between the two dispersive elements.

\begin{figure}[htbp]
\begin{center}
\psfrag{h}[c][c][0.8]{\shortstack{incident wave\\$\psi(x,y, z; t)$}}
\psfrag{g}[c][c][0.95]{$z=0$}
\psfrag{a}[c][c][0.8]{\shortstack{spatio-temporal\\metasurface $t_\text{m}(x,y)$}}
\psfrag{d}[c][c][0.8]{\shortstack{output plane \\$[x_0(\nu),\;y_0(\nu)]$}}
\psfrag{e}[c][c][0.95]{$\Delta z = f$}
\psfrag{x}[c][c][0.95]{$x$}
\psfrag{y}[c][c][0.95]{$y$}
\psfrag{z}[c][c][0.95]{$z$}
\psfrag{v}[c][c][0.95]{$\nu$}
\psfrag{r}[l][c][0.85]{$\psi(x,y; 0_+)$}
\psfrag{u}[r][c][0.85]{$\psi(x,y; f)$}
\includegraphics[width=0.7\columnwidth]{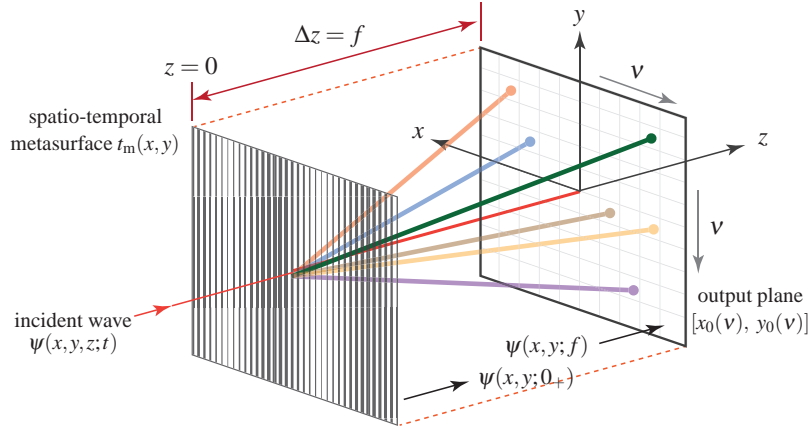}
\caption{Spatio-temporal metasurface decomposing the temporal spectrum of a broadband wave, $\psi(x,y,z;t)$, in two spatial dimensions, where each frequency $\nu$ is mapped onto a specific point $[x_0(\nu),y_0(\nu)]$ in the focal plane at $z=f$.}
\label{Fig:Metasurface2D}
\end{center}
\end{figure}

One may naturally ask the following question: can such a spectral shower be obtained using a \emph{single dispersive element}, so that the system size reduces to $\Delta z = f$, as suggested in Fig.~\ref{Fig:Metasurface2D}? This question is addressed here using a metasurface as the dispersive element that will map the temporal frequencies, $\nu$, of $\psi(x,y,0_-, t)$ onto the spatial points $[x_0(\nu), y_0(\nu)]$ in the output plane $z=f$. The sought metasurface will be characterized by a complex transmittance function directly operating on the input wave and producing the spectrally resolved output wave at $z=f$.

\subsection{An Equivalent Free-space System}

\begin{figure}[htbp]
\begin{center}
\psfrag{x}[c][c][0.95]{$x$}
\psfrag{y}[c][c][0.95]{$y$}
\psfrag{z}[c][c][0.95]{$z$}
\psfrag{f}[c][c][0.85]{$z=f$}
\psfrag{g}[c][c][0.95]{$\lambda = 1/\nu$}
\psfrag{b}[c][c][0.65]{$\delta\approx 0$}
\psfrag{c}[c][c][0.85]{$z=0$}
\psfrag{d}[c][c][0.85]{\shortstack{output \\plane}}
\psfrag{e}[c][c][0.85]{\shortstack{incident wave\\$\psi(x,y;0_-)$}}
\psfrag{w}[c][c][0.8]{$|t_g(y)|$}
\psfrag{m}[c][c][0.85]{$m$}
\psfrag{p}[c][c][0.85]{$d(x)$}
\psfrag{n}[c][c][0.85]{$\Lambda$}
\psfrag{q}[c][c][0.85]{\shortstack{thin dispersive\\ wedge $n(\lambda)$}}
\psfrag{r}[c][c][0.7]{$\psi(x,y, l_+)$}
\psfrag{s}[c][c][0.7]{$\psi(x,y, 0_-) $}
\psfrag{t}[c][c][0.7]{$+$}
\psfrag{u}[c][c][0.7]{$\psi(x,y, f) $}
\psfrag{v}[c][c][0.85]{$=$}
\psfrag{a}[c][c][0.85]{\color{red}\shortstack{Equivalent Metasurface \\ $t_m(x,y)$} }
\psfrag{A}[c][c][0.75]{\shortstack{thin wedge\\ $t_w(x,y)$}}
\psfrag{B}[c][c][0.75]{\shortstack{diffraction \\grating $t_g(x,y)$}}
\psfrag{C}[c][c][0.75]{\shortstack{focussing \\lens $t_l(x, y)$}}
\psfrag{s}[c][c][0.7]{$[x_0(\nu),\;y_0(\nu)]$}
\includegraphics[width=\columnwidth]{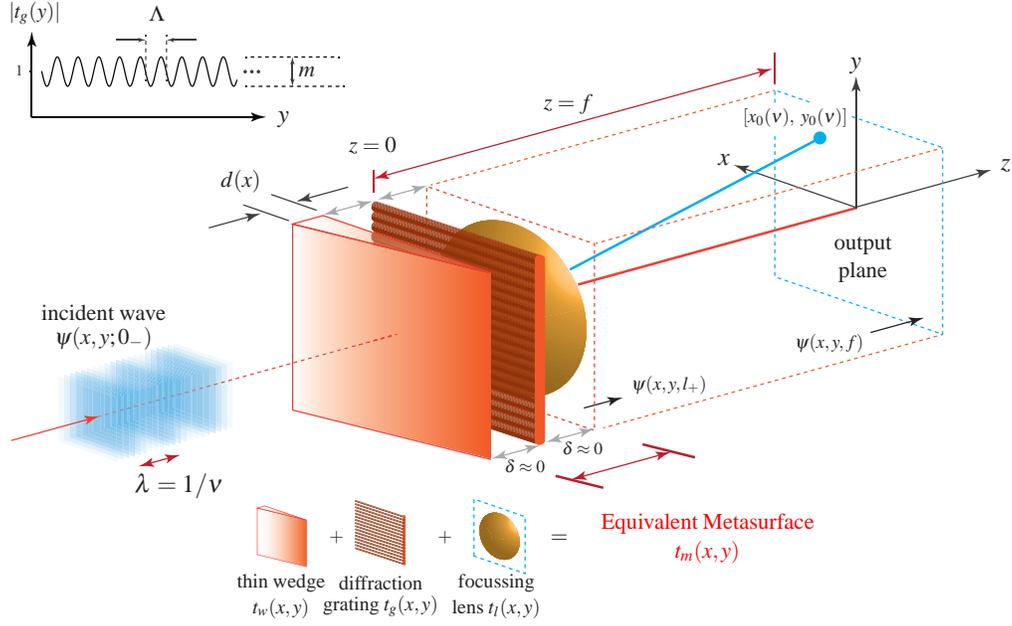}
\caption{Equivalent free-space system for mapping temporal frequencies $\nu$ onto spatial points $[x_0(\nu),y_0(\nu)]$.}
\label{Fig:Setup}
\end{center}
\end{figure}

The complex transmittance function, $t_\text{m}(x,y;\nu)$, of the spatio-temporal metasurface can be derived by first considering the equivalent free-space system of Fig.~\ref{Fig:Setup}. This system consists of the following three cascaded elements: 1)~a thin linear wedge of transmittance $t_\text{w}(x,y)$ with a wavelength dependent refractive index $n(\nu)$, 2)~a sinusoidal diffraction grating of transmittance $t_\text{g}(x,y)$, and 3)~a focussing lens of transmittance $t_\text{l}(x,y)$ with focal length $f$. The thin wedge and the lens are directly stacked with the diffraction grating, and refraction within the wedge is assumed to be negligible. In addition, paraxial wave propagation is assumed for simplicity. It will be shown that this system is equivalent to the spatio-temporal metasurface of Fig.~\ref{Fig:Metasurface2D}.

Let us first consider the wedge, whose thickness may be written $d(x) = a_0 (x + w_0)$, where $a_0$ and $w_0$ are constants. The corresponding transmittance is

\begin{equation}
t_\text{w}(x,y;\nu)
=\exp[-jkn(\nu)d(x)]
=\underbrace{\exp[-jkn(\nu)a_0w_0]}_{t_0(\nu)}\exp[-jkn(\nu)a_0x],
\end{equation}

\noindent while the complex transmittance function of the diffraction grating is given by

\begin{equation}
t_\text{g}(x,y)
=1+m\cos(2\pi f_0y),
\end{equation}

\noindent $f_0$ and $\Lambda=1/f_0$ are the grating frequency and period, respectively.

Consider a plane-wave propagating along the $z-$direction,

\begin{equation}
\psi(x,y,z;t) = \text{Re}\{e^{-j(kz - \nu t)}\},
\end{equation}

\noindent so that $\psi(x,y,0_-;t) =\psi(x,y,0_-)  = 1$, where the explicit time dependence $e^{j\nu t}$ was dropped again. The wave just after the grating (and before the lens) is then obtained as

\begin{equation}
\begin{split}
\psi(x,y, 0_+)
&=t_\text{w}(x,y; \nu)t_\text{g}(x,y; \nu)\psi(x,y, 0_-)
=t_0(\nu)\exp[-jkn(\nu)a_0x]\left[1+m\cos(2\pi f_0y)\right]\\
&=t_0(\nu)\left\{\underbrace{\exp[-jkn(\nu)a_0x]}_\text{$0^\text{th}$ order}+m\underbrace{\exp[-jkn(\nu)a_0x]\cos(2\pi f_0y)}_\text{$\pm1^\text{st}$ order}\right\}.\\
\end{split}
\end{equation}

\noindent The first term in the last relation corresponds to the $0^\text{th}$ diffraction order of the system, whose location in the output plane is independent of $\nu$ along the $y-$direction of the grating~\cite{Goodman_Fourier_Optics}, and will therefore be ignored. Considering the lens transmittance function \mbox{$t_\text{l}(x,y; \nu)=\exp[-j\frac{\pi\nu}{ fc}(x^2 + y^2)]$}, the wave at the output of the lens is obtained as

\begin{equation}
\psi(x,y, l_+)=mt_0(\nu)\exp[-jkn(\nu)a_0x]\cos(2\pi f_0y)\exp\left[-j\frac{\pi\nu}{ fc}(x^2 + y^2)\right],\label{Eq:LensOutput}
\end{equation}

\noindent which leads to the field intensity (App.~\ref{Sec:OP})

\begin{equation}
I(x,y,f)= |\psi(x,y, f)|^2=m^2\nu^2\frac{\left|\delta\{x+a_0n(\nu)f\}\delta(y+ f_0 c f/\nu)\right|}{2( fc)^2},\label{Eq:OutputIntensity}
\end{equation}

\noindent where only one of the diffraction orders is kept, since the other one is its symmetric counterpart. This equation corresponds to the sought frequency-to-space mapping law of the system, where each frequency $\nu$ is mapped onto a specific point $(x_0, y_0)$ on the output plane according to

\begin{subequations}
\begin{equation}
x_0(\nu) = a_0n(\nu)f,
\end{equation}
\begin{equation}
y_0(\nu) =  - f_0\left(\frac{c}{\nu}\right) f.
\end{equation}\label{Eq:Scanning_Law}
\end{subequations}

Equation~(\ref{Eq:Scanning_Law}b) is the conventional 1-D scanning relation along the $y-$direction provided by a diffraction grating. On the other hand, frequency scanning along the $x-$direction follows from the \emph{frequency dependent refractive index, $n(\nu)$, of the wedge}. Leveraging the two mechanism offers precise control over the $x-$ and $y-$direction decompositions and hence enables efficient 2-D spectral-to-spatial mapping.

The resolution enhancement factor of the dispersive wedge may be defined as (App.~\ref{Sec:Resolution})

\begin{equation}
R = \frac{\int_{\lambda_1}^{\lambda_2}\left[1+\left(\frac{a_0n^{\prime}}{f_0}\right)^2 \right]^{1/2}d\lambda}{\lambda_2 - \lambda_1},\label{Eq:resolution}
\end{equation}

\noindent where $n^{\prime}(\lambda) = dn(\lambda)/d\lambda$. In this relation, the impact of the wedge temporal dispersion is explicit: The enhancement is proportional to the dispersion parameter, $n^{\prime}(\lambda)$. For a non-dispersive wedge, $R=1$.

\subsection{Metasurface Transmittance Function}

The system of Fig.~\ref{Fig:Setup} can be replaced by a single spatio-temporal metasurface with complex transmittance function

\begin{equation}
t_\text{m}(x, y; \nu) = \overbrace{\exp[-jka_0n(\nu)x]}^\text{Dispersive wedge}\underbrace{\exp\left[-j\frac{\pi\nu}{fc}(x^2 + y^2)\right]}_\text{focussing lens}\overbrace{\cos(2\pi f_0y)}^\text{diffraction grating},
\end{equation}

\noindent combining the effects of the dispersive wedge, diffraction grating and focussing lens in a single element. This metasurface will focus the spectral contents of an incident wave following the frequency-to-space mapping given by (\ref{Eq:Scanning_Law}) at the focal plane, located at $z=f$, by directly operating on the incident field as

\begin{subequations}
\begin{equation}
\psi(x,y,f;\nu) = \underbrace{\mathcal{F}_t\left[\psi(x,y,0_-;t) \right]t_\text{m}(x, y; \nu)}_{\text{metasurface output}}\overbrace{\ast h(x,y)},
\end{equation}
\begin{equation}
\psi(x,y,f;t) = \mathcal{F}_t^{-1}\{\psi(x,y,f;\nu) \},
\end{equation}\label{Eq:Model}
\end{subequations}

\noindent where $\mathcal{F}_t(\cdot)$ and $\mathcal{F}_t^{-1}(\cdot)$ are the temporal Fourier transform and inverse Fourier transform operators, respectively, and where the convolution product is naturally with respect to space with

\begin{equation}\label{Eq:FS_imp_resp}
h(x,y)=\frac{e^{jkz}}{j\lambda z}\exp\left[j\frac{\pi}{\lambda z}(x^2 + y^2)\right]
\end{equation}

\noindent as the free-space impulse response under paraxial approximation~\cite{Goodman_Fourier_Optics}~(4-21). 

This completes the synthesis of the spatio-temporal metasurface for real-time 2-D spectrum analysis.
	
\section{Results}

\subsection{Lorentz Dispersive Wedge}

\begin{figure}[htbp]
\begin{center}
\psfrag{A}[c][c][0.95]{frequency, $\nu$ (THz)}
\psfrag{B}[c][c][0.95]{susceptibility $\chi$}
\psfrag{C}[l][c][0.95]{$\chi^\prime$}
\psfrag{D}[l][c][0.95]{$\chi^{\prime\prime}$}
\psfrag{c}[c][c][0.95]{$\nu_r$}
\psfrag{d}[c][c][0.95]{$\nu_{1, r}$}
\psfrag{e}[c][c][0.95]{$\nu_{2,r}$}
\psfrag{f}[c][c][0.95]{$\nu_{3,r}$}
\includegraphics[width=0.75\columnwidth]{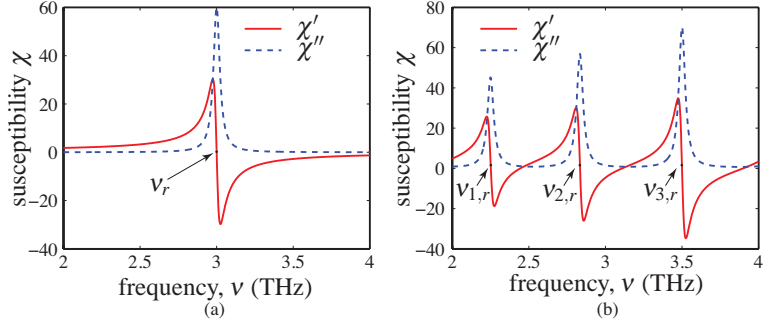}
\caption{Lorentz susceptibility responses for a) single $(M=1)$ resonance, and b) multiple ($M=3$) resonances. $\Delta\nu = 0.05$~THz.}
\label{Fig:LorentzDispersion}
\end{center}
\end{figure}

Consider a wedge which exhibits Lorentz dispersion with multiple resonances at frequencies $\nu_{i, r}$ and line width $\Delta\nu$, $i\in[1,M]$. The corresponding susceptibility is given by

\begin{equation}
\chi(\nu) = \sum_{i=1}^M \frac{\nu_{i, r}^2}{(\nu_{i, r}^2 - \nu^2) + j\nu\Delta\nu},\label{Eq:Lorentz}
\end{equation}

\noindent and the corresponding refractive index is $n(\nu) = \sqrt{1 + \chi(\nu)}$. Figure~\ref{Fig:LorentzDispersion} plots the real and imaginary parts of the susceptibilities for typical Lorentz dispersion responses in the cases of single resonance ($M=1$) and triple resonance ($M=3$). Consistently with Kramers-Kronig relations for an electromagnetic medium~\cite{Jackson_book_CED}, resonances coincide with maximal absorption. Moreover, it may be observed that for given line width, the susceptibility swing increases with increasing resonant frequencies. Engineering the locations of the resonant frequencies will thus allow one to tailor the dispersion response of the medium over a broad frequency range.

\subsection{Results}

\begin{figure}
\begin{center}
\begin{minipage}[c]{.5\columnwidth}
\vspace{0pt}
\centering
  \psfrag{A}[c][c][0.95]{distance $x/\lambda_0$}
\psfrag{B}[c][c][0.95]{distance $y/\lambda_0$}
\psfrag{C}[c][c][0.95]{$|\psi(x,y, 0_+)|^2$}
\includegraphics[width=0.7\columnwidth]{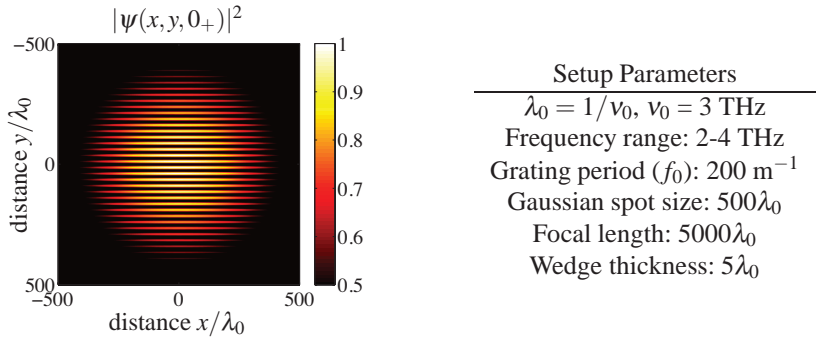}
\end{minipage}
\begin{minipage}[c]{.4\columnwidth}
\vspace{0pt}
\centering
\begin{tabular}[t]{c}
Setup Parameters\\
 \hline
    $\lambda_0 = 1/\nu_0$, $\nu_0$ = 3~THz \\
        Frequency range: 2-4~THz\\
        Grating period ($f_0$): 200~m$^{-1}$\\
    Gaussian spot size: $500\lambda_0$\\
    Focal length: $5000\lambda_0$\\
    Wedge thickness: $5\lambda_0$
  \end{tabular}
\end{minipage}
\caption{Normalized wave intensity $|I(x,y,0_+)|$ corresponding to Gaussian beam illumination of the wedged diffraction grating (left), and parameters used in the numerical examples (right). The modulation index of the grating is assumed to be $m=1$. }\label{Fig:Setup_Parameters}
\end{center}
\end{figure}

Let us assume Gaussian beam illumination in the system of Fig.~\ref{Fig:Metasurface2D}. In the forthcoming numerical examples, the previous paraxial propagation constraint is suppressed and an exact transfer function is used in~\eqref{Eq:Model}, i.e. $\tilde{H}(k_x, k_y) = \exp[-j(k_x^2 + k_y^2)^{1/2}z]$~\cite{Goodman_Fourier_Optics}. The field distribution just after the wedged grating is shown in Fig.~\ref{Fig:Setup_Parameters} along with system parameters. The choice of these parameters is somewhat arbitrary but corresponds to quantities that can be easily realized in current technologies.

\begin{figure}
\begin{center}
\psfrag{A}[c][c][0.95]{distance $x/\lambda_0$}
\psfrag{B}[c][c][0.95]{distance $y/\lambda_0$}
\psfrag{C}[c][c][0.95]{\color{white}$-1^\text{st}$-order}
\psfrag{E}[c][c][0.95]{\color{white}$+1^\text{st}$-order}
\psfrag{D}[c][c][0.8]{\color{white}$\boxed{a_0=0}$}
\psfrag{F}[c][c][0.95]{\color{white}$\nu$}
\psfrag{G}[l][c][0.95]{\color{white}$(a_0nf)$}
\psfrag{H}[c][c][0.8]{\color{white}$\boxed{a_0=0.01}$}
\psfrag{J}[c][c][0.95]{$20\log|\psi(x,y, f)|$ (dB)}
\includegraphics[width=0.9\columnwidth]{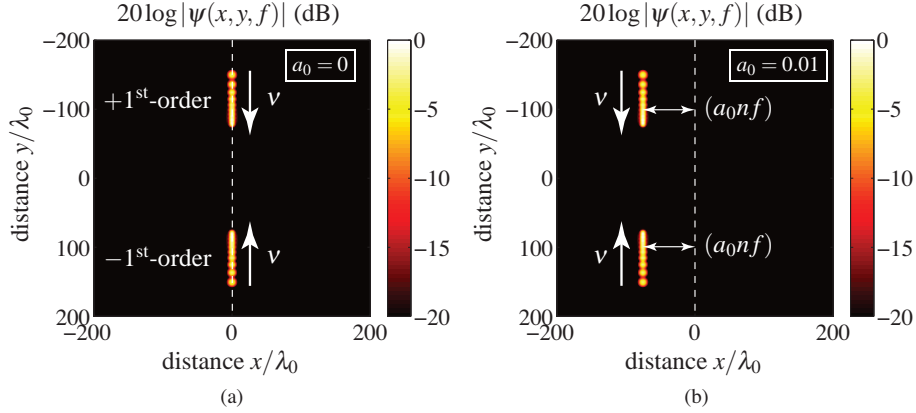}
\caption{Normalized wave intensity $|I(x,y,f)|$ computed with~\eqref{Eq:Model} showing~1-D frequency scanning in the output plane for uniformly spaced frequencies within the specified bandwidth a)~in the absence of wedge, and b)~in the presence of a non-dispersive wedge of refractive index $n = 3$. Only one of the two diffraction orders are shown. Each bright spot represents a distinct frequency $\nu$.}\label{Fig:NonDisp_Wedge}
\end{center}
\end{figure}

First, consider a trivial case where no wedge is present. The system behaves then as a conventional 1-D frequency scanned system where the diffraction grating performs conventional spectral decomposition along 1-dimension ($y-$direction here). Figure~\ref{Fig:NonDisp_Wedge}(a) shows the intensity distribution in the focal plane for several frequencies covering the specified bandwidth. The high intensity beam spots move towards the centre along the $y-$direction following the scanning law~(\ref{Eq:Scanning_Law}b) as frequency increases. Consider next the case where a non-dispersive wedge with with $n(\nu) = n_0$ is introduced. Now scanning is shifted along the $x-$direction to $x_0 = a_0n_0f$ following the scanning law~(\ref{Eq:Scanning_Law}a), as shown in Fig.~\ref{Fig:NonDisp_Wedge}(b). This is still a 1-D scan.

\begin{figure}[htbp]
\begin{center}
\subfigure[]{
\psfrag{A}[c][c][0.95]{distance $x/\lambda_0$}
\psfrag{B}[c][c][0.95]{distance $y/\lambda_0$}
\psfrag{C}[l][c][0.95]{\color{white}$a_0fn(\nu)$}
\psfrag{F}[c][c][0.95]{\color{white}$\nu$}
\psfrag{D}[c][c][0.8]{\color{white}$\boxed{a_0=0.005}$}
\psfrag{F}[c][c][0.95]{\color{white}$\nu$}
\psfrag{E}[c][c][0.7]{$\Delta x = a_0(n_\text{max}- n_\text{min})f$}
\psfrag{G}[c][c][0.7]{$\Delta y = f_0(\lambda_\text{max}- \lambda_\text{min})f$}
\psfrag{H}[c][c][0.95]{frequency $\nu$ (THz)}
\psfrag{J}[c][c][0.95]{$20\log|\psi(x,y, f)|$ (dB)}
\includegraphics[width=0.8\columnwidth]{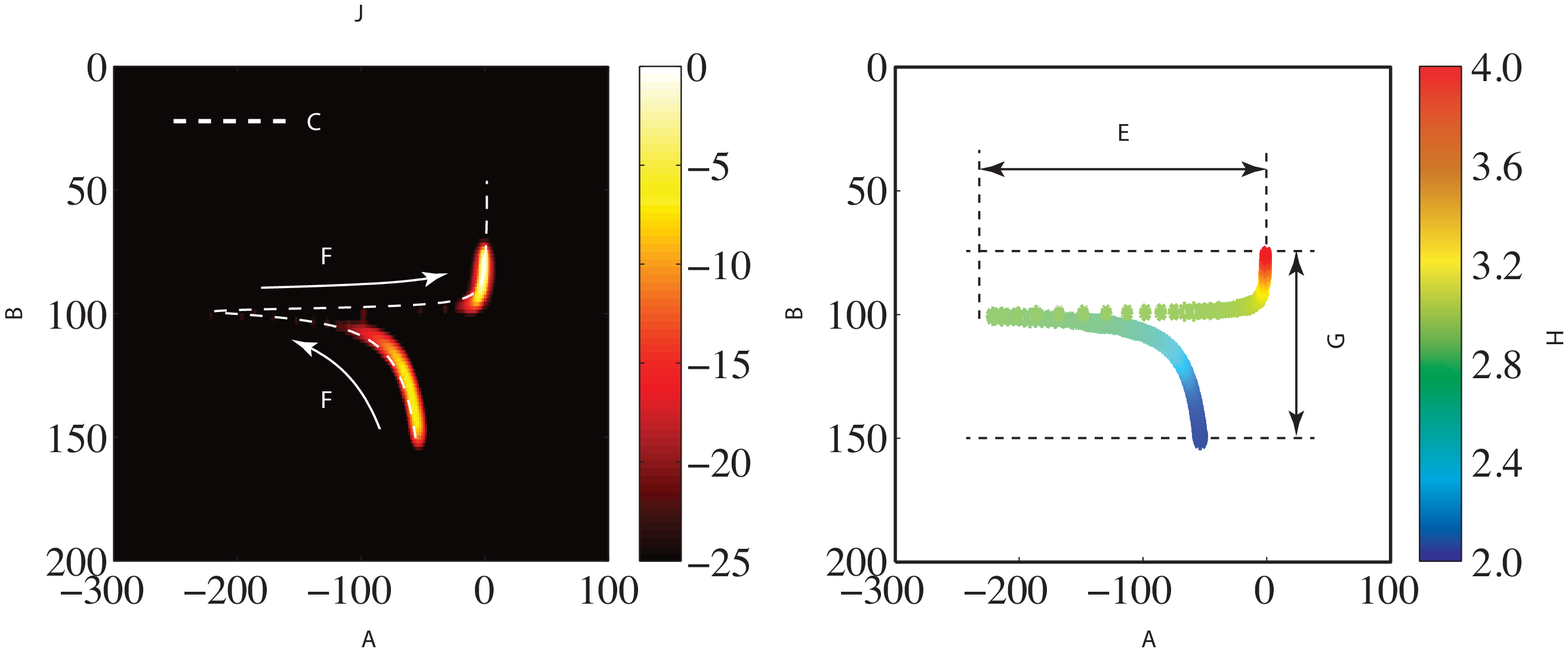}}
\subfigure[]{
\psfrag{A}[c][c][0.95]{distance $x/\lambda_0$}
\psfrag{B}[c][c][0.95]{distance $y/\lambda_0$}
\psfrag{C}[l][c][0.95]{\color{white}$a_0fn(\nu)$}
\psfrag{F}[c][c][0.95]{\color{white}$\nu$}
\psfrag{D}[c][c][0.8]{\color{white}$\boxed{a_0=0.005}$}
\psfrag{F}[c][c][0.95]{\color{white}$\nu$}
\psfrag{E}[c][c][0.7]{$\Delta x = a_0(n_\text{max}- n_\text{min})f$}
\psfrag{G}[c][c][0.7]{$\Delta x = f_0(\lambda_\text{max}- \lambda_\text{min})f$}
\psfrag{H}[c][c][0.95]{frequency $\nu$ (THz)}
\psfrag{J}[c][c][0.95]{$20\log|\psi(x,y, f)|$ (dB)}
\includegraphics[width=0.8\columnwidth]{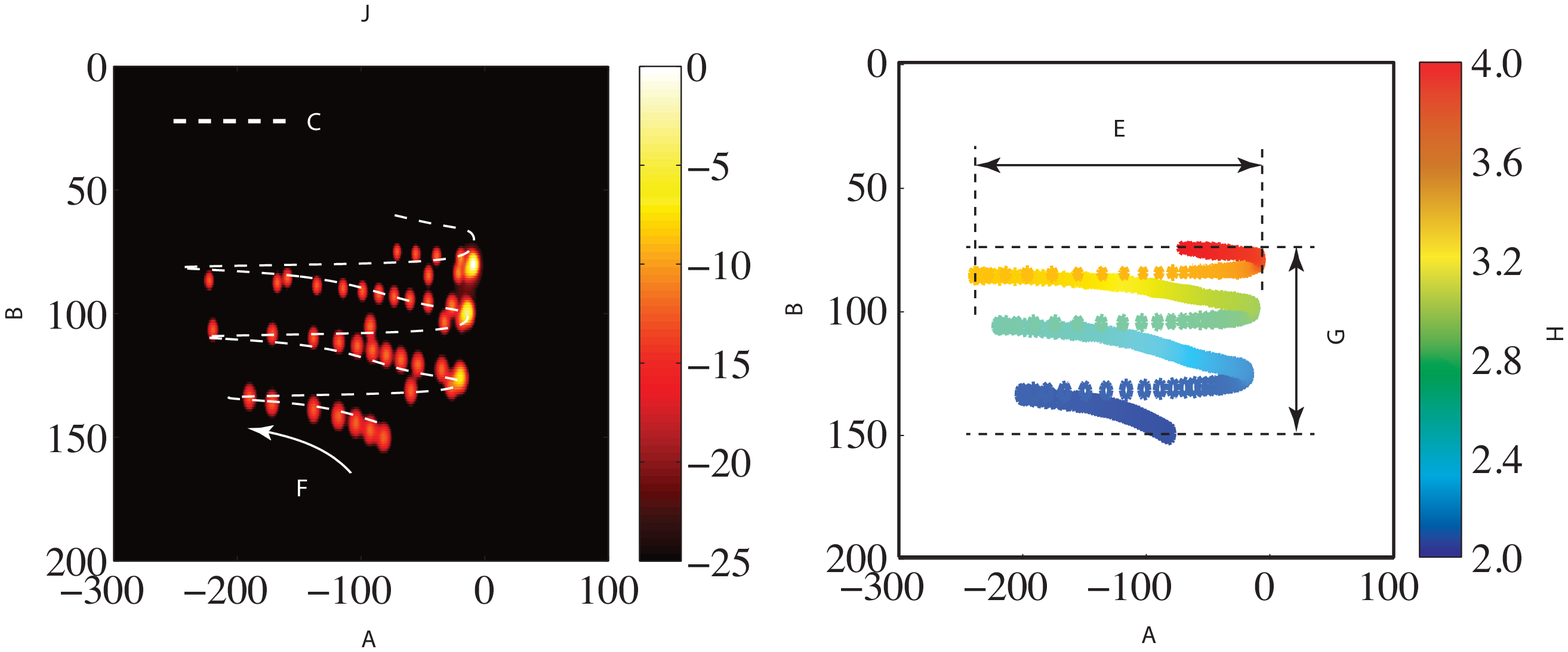}}
\caption{Normalized wave intensity $|I(x,y,f)|$ computed with~(\ref{Eq:Model}) showing 2-D frequency scanning in the output plane for uniformly spaced frequencies within the specified bandwidth in the presence of a dispersive wedge exhibiting a)~single resonance, as in Fig.~\ref{Fig:LorentzDispersion}(a), and b)~multiple resonances, as in Fig.~\ref{Fig:LorentzDispersion}(b). Left graphs show the superimposed beam spots of different frequencies while right graphs show the corresponding 3-dB contour plots.}
\label{Fig:Disp_Grating}
\end{center}
\end{figure}

The case of real interest is that where the wedge exhibits the dispersive response plotted in Fig.~\ref{Fig:LorentzDispersion}. Figure~\ref{Fig:Disp_Grating} shows the corresponding output intensity distribution for a single-resonant dispersive response and a multi-resonant dispersion response. As expected, the frequency spots are now moving along both the $x-$ and $y-$directions as frequency is varied, following the scanning law~(\ref{Eq:Scanning_Law}). The 3-dB contour plots show the exact trajectory of frequencies in the focal plane. The frequency scanning closely follows the Lorentz dispersion response of the wedge. The frequency scanning region is $\Delta x = a_0(n_\text{max}- n_\text{min})f$ and $\Delta y = f_0(\lambda_\text{max}- \lambda_\text{min})f$. The number of spatial scanning periods is equal to the number of resonant points $M$ in the Lorentz response. This result successfully demonstrates the principle of frequency scanning in 2-dimensions using a dispersive wedged-grating.

\section{Conclusion}

A spatio-temporal metasurface has been proposed to spectrally decompose electromagnetic waves in two spatial dimensions, and this metasurface has been demonstrated using Fourier optics. The characteristics of the metasurface have been derived using an equivalent free-space system consisting of a frequency dispersive wedge cascaded with a conventional diffraction grating and a focusing lens. It has been demonstrated that to achieve 2-D frequency scanning in space, the metasurface needs to exhibit a multi-resonant response in a broad frequency range, in conjunction with the conventional 1-D scanning property of a diffraction grating. The size of the proposed system corresponds to the focal distance specified in the phase profile of the metasurface, and the frequency resolution increases with the number of resonance in the dispersive response of the wedge. Multi-resonant structures have been demonstrated in various systems, using multiple-plasmon resonances~\cite{Multi_Plasmon_Resonance,MultiFanoResonance_MS,broadband_MS} and unit-cell designs with multiple resonating elements characterize by different shapes and sizes~\cite{Multiresonant_Nanoantenna,Superstarte_Resonant_MS}, and it is therefore expected that multi-resonance could be easily achievable in the proposed metasurface. This metasurface may contribute to pave the road for next-generation light-weight, compact and super-resolution spectrum analyzers.

\section{Appendix}

\subsection{Derivation of Output Intensity $I(x,y,f)$ (\ref{Eq:OutputIntensity})}\label{Sec:OP}

The output wave, in the focal plane, is 

\begin{equation}
\psi(x,y,f)=\psi(x,y,l_+)\ast h(x,y).\label{Eq:FocalOP}
\end{equation}

\noindent Substituting~\eqref{Eq:LensOutput} and~\eqref{Eq:FS_imp_resp} into this expression, and using Euler identity $\cos(2\pi f_0Y) = [\exp(j2\pi f_0Y) + \exp(-j2\pi f_0Y)]/2$, yields for the intensity of the output wave

\begin{equation}
\begin{split}
&|I(x,y;z =f)| = \left|\psi(x,y;l_+)\ast h(x, y)\right|^2\\
& = \left|mt_0(\lambda)\exp[-jkn(\lambda)a_0x]\cos(2\pi f_0y)\exp\left[-j\frac{\pi}{\lambda f}(x^2 + y^2)\right] \ast \frac{e^{jkf}}{j\lambda f}\exp\left[j\frac{\pi}{\lambda f}(x^2 + y^2)\right] \right|^2\\
& = \left(\frac{mt_0(\lambda)}{\lambda f}\right)^2\left|\int\exp[-jkn(\lambda)a_0X]\exp\left(-j\frac{2\pi x}{\lambda f}X\right)dX\int\cos(2\pi f_0Y) \exp\left(-j\frac{2\pi y}{\lambda f}Y\right) dY\right|^2\\
& = m^2t_0^2(\lambda)\frac{\left|\delta\{x+a_0n(\lambda)f\}\right|}{(\lambda f)^2}\left[\frac{\delta(y+ f_0 \lambda f) + \delta(y- f_0 \lambda f)}{2}\right].
\end{split}
\end{equation}

\subsection{Derivation of the Resolution Enhancement Factor $R$ (\ref{Eq:resolution})}\label{Sec:Resolution}

\begin{figure}
\begin{center}
\psfrag{A}[c][c][0.95]{$y_0(\nu)$}
\psfrag{B}[c][c][0.95]{$x_0(\nu)$}
\psfrag{C}[c][c][0.95]{$\nu_\text{min}$}
\psfrag{D}[c][c][0.95]{$\nu_\text{max}$}
\psfrag{E}[c][c][0.95]{$\Delta\ell$}
\psfrag{F}[c][c][0.95]{$dy_0$}
\psfrag{G}[c][c][0.95]{$dx_0$}
\psfrag{H}[c][c][0.95]{$\mathbf{dr}$}
\psfrag{I}[c][c][0.95]{$\Delta r_\text{d}$}
\includegraphics[width=0.4\columnwidth]{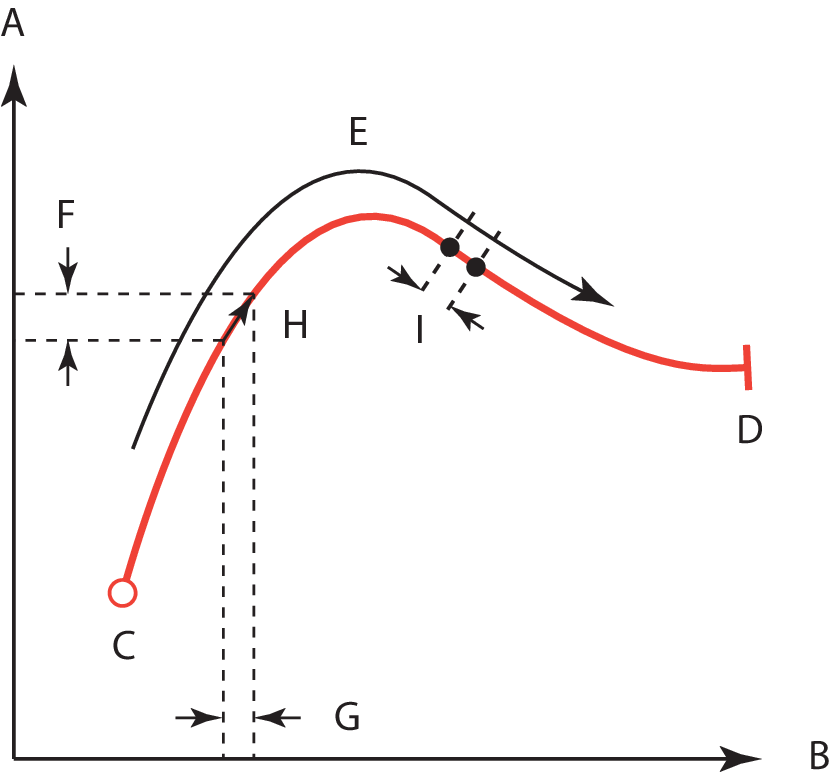}
\caption{Total path length, $\Delta\ell$, and detector resolution, $\Delta r_\text{d}$, in the output plane ($z=f$) for computation of the resolution enhancement factor \eqref{Eq:res_enhanc_fact}.}\label{Fig:Path}
\end{center}
\end{figure}

The frequency resolution in the output plane (Fig.~\ref{Fig:Setup}), for a given bandwidth \mbox{$\Delta\nu = \nu_\text{max} - \nu_\text{min}$}, depends on the total path length, $\Delta\ell$, and on the spatial resolution of the detector, $\Delta r_\text{d}$, indicated in Fig.~\ref{Fig:Path}. If $\Delta r_\text{d}$ and $\Delta\nu$ are fixed, the frequency resolution directly depends on the total path $\Delta\ell$. Using the scanning law of~\eqref{Eq:Scanning_Law}, the total path length is found to be

\begin{subequations}
\begin{equation}
\Delta \ell = \int|\mathbf{dr}| = \int (dx_0^2 + dy_0^2)^{1/2} = ff_0\int_{\lambda_1}^{\lambda_2}\left[1+\left(\frac{a_0}{f_0}\right)^2n^{\prime2} \right]^{1/2}d\lambda,
\end{equation}
\noindent after using
\begin{equation}
\frac{dx_0}{d\lambda} =  a_0f\frac{dn(\lambda)}{d\lambda} =  a_0fn^\prime(\lambda)\quad\text{and}\quad \frac{dy_0}{d\lambda} =  -f_0f.
\end{equation}
\end{subequations}

\noindent The resolution enhancement factor $R$ in the dispersive wedge system compared to the system without the wedge is finally defined as

\begin{equation}\label{Eq:res_enhanc_fact}
R = \frac{\int_{\lambda_1}^{\lambda_2}\left[1+\left(\frac{a_0n^{\prime}}{f_0}\right)^2 \right]^{1/2}d\lambda}{\int_{\lambda_1}^{\lambda_2}d\lambda}= \frac{\int_{\lambda_1}^{\lambda_2}\left[1+\left(\frac{a_0n^{\prime}}{f_0}\right)^2 \right]^{1/2}d\lambda}{\lambda_2 - \lambda_1}.
\end{equation}

\end{document}